# Chiral magnetoresistance in the Weyl semimetal NbP

*Anna Corinna Niemann,[1,2]\* Johannes Gooth,[1,3]\* Shu-Chun Wu,[4] Svenja Bäßler,[1] Philip Sergelius,[1] Ruben Hühne,[2] Bernd Rellinghaus,[2] Chandra Shekhar,[4] Vicky Süß,[4] Marcus Schmidt,[4] Claudia Felser,[4] Binghai Yan[4,5] and Kornelius Nielsch[1,2]*

[1]Institute of Nanostructure and Solid State Physics, Universität Hamburg, Jungiusstraße 11, 20355 Hamburg, Germany

[2] Leibniz Institute for Solid State and Materials Research Dresden, Institute for Metallic Materials, Helmholtzstraße 20, 01069 Dresden, Germany

[3]IBM Research-Zurich, Säumerstrasse 4, 8803 Rüschlikon, Switzerland

[4]Max Planck Institute for Chemical Physics of Solids, Nöthnitzer Straße 40, 01187 Dresden, Germany

[5]Max Planck Institute for Physics of Complex Systems, Nöthnitzer Straße 38, 01187 Dresden, Germany

*These authors contributed equally to this work.

**NbP is a recently realized Weyl semimetal (WSM), hosting Weyl points through which conduction and valence bands cross linearly in the bulk and exotic Fermi arcs appear. However, the most intriguing transport phenomenon of a WSM, the chiral anomaly-induced negative magnetoresistance (NMR) in parallel electric and magnetic fields, has yet to be observed in NbP. In intrinsic NbP the Weyl points lie far from the Fermi energy, making chiral magneto-transport elusive. Here, we use Ga-doping to relocate the Fermi energy in NbP sufficiently close to the Weyl points, for which the different Fermi surfaces are verified by resultant quantum oscillations. Consequently, we observe a NMR for parallel electric and magnetic fields, which is considered as a signature of the chiral**

**anomaly in condensed-matter physics. The NMR survives up to room temperature, making NbP a versatile material platform for the development of Weyltronic applications.**

Weyl semimetals[1,2,3] are a recently realized, new topological matter[4,5,6] in which conduction and valence bands touch linearly near the Fermi energy ($E_F$). Weyl states are closely related to the well-known Dirac states,[7,8] where both time reversal and inversion symmetries are preserved. In contrast, Weyl states emerge when one of these symmetries is broken. The band touching points in a Weyl semimetal – so-called Weyl nodes – always come in spatially separated pairs of opposite chirality ($\chi = \pm 1$), which distinguish them from Dirac semimetals having two degenerate Weyl nodes of opposite chirality that form one Dirac node. Therefore, exotic features of Weyl semimetals include Fermi arcs, which connect two Weyl nodes of opposite chirality. This occurs even in the absence of an external magnetic field, which would be required for the realization of Fermi arcs in Dirac semimetals.

For relativistic Weyl and Dirac fermions, chirality is, in principle, a strictly conserved quantum number, which gives the direction of the fermion spin relative to the direction of its linear momentum. Nevertheless, when these relativistic fermions are subjected to electromagnetic fields with parallel electric and magnetic field components ($E \| B$), the gauge invariance also has to be taken into account, and a breakdown of chiral symmetry occurs. The physical origin of this phenomenon lies in the splitting of the band structure into Landau levels in the presence of a magnetic field. In this scenario, only the zeroth Landau level of the Weyl band exhibits uni-directional fermion velocity, forward or backward along the magnetic field direction depending on its chirality. Without an additional electric field, the forward and backward moving zeroth Landau levels are equally filled; however, applying $E \| B$ leads to chiral charge pumping between the two branches, with a flow rate that is proportional to $E \cdot B$. The electrons are forced to move – for example – in the forward direction, and consequently, the forward moving zeroth Landau level is filled more than the backward moving one (**Fig. 3 a**). Such breaking of the chiral symmetry is referred to as the Adler-Bell-Jackiw or chiral anomaly.[9,10] In electrical transport measurements this anomaly leads to an additional topological current in parallel to the aligned electric and magnetic fields, decreasing the resistance of a Weyl semimetal with increasing magnetic field. Such a negative magnetoresistance (NMR) is considered to be a strong signature of broken chiral symmetry in condensed matter systems.[11,12] In the limit of weak $B$, Son and Spivak showed, in fact, that the resistance - in a condensed matter system with broken chiral symmetry and $E \| B$ - is inverse proportional to the squared magnetic field ($R \sim 1/B^2$).[13] NMR related to broken chirality has already been reported for Dirac semimetals such as $Cd_2As_3$ and $NaBi_3$ and the Weyl metal TaAs.[14,15,16]

For TaAs-family compounds, which are recently discovered inversion-breaking Weyl semimetals,[4, 5, 17, 18, 19] two groups of Weyl points were found and classified into four pairs in the $k_z = 0$ plane (called W1) and eight pairs off the $k_z = 0$ plane (called W2) in the first Brillouin zone.[20] NbP is the lightest member of the TaAs family. It has a non-centrosymmetric crystal structure in a tetragonal lattice (space group $I4_1md$), as sketched in **Fig. 1 b**. Intrinsic NbP has been intensively studied by magneto-transport[20] and quantum oscillation measurements.[21,22] Ultrahigh mobility ($5 \cdot 10^6$ cm$^2$ V$^{-1}$ s$^{-1}$ at 9 T and 1.85 K)[20] and huge magnetoresistance ($8.5 \cdot 10^5$ % at 9 T and 1.85 K) attributed to electron-hole resonance[23] have been observed.[20] Quantum oscillation measurements combined with *ab initio* calculations revealed that the W1 and W2 Weyl nodes are located -57 meV and +5 meV away from the intrinsic Fermi level ($E_{F0}$) of the pristine NbP sample without intentional doping.[21]

For the magneto-transport measurements, elongated, Ga-doped NbP micro-ribbons of 50 μm x 2.46 μm x 526 nm were prepared by Ga focused ion beam (FIB) etching into the bulk sample. We used X-ray diffraction (XRD) measurements on our single crystalline NbP bulk sample to identify the crystal axis of the NbP micro-ribbon. Additionally, the XRD pattern of the (001) surface of the NbP bulk sample, as given in **Fig. 1 c**, shows the good quality of the sample with the full-width half-maximum being no larger than 0.15 degrees. The NbP micro-ribbon was extracted from the bulk sample using a micromanipulator and electrically contacted using laser lithography, followed by a Ti/Pt metallization process. A micrograph of the electrically contacted sample can be seen in **Fig. 1 d**. The high aspect ratio of the micro-ribbon with contact lines across the full width of the sample is chosen to supress current jetting and to ensure a homogenous *E*-field distribution, respectively.[24] The Ga concentration on the surface of the micro-ribbon (analysed by scanning electron microscopy energy dispersive x-ray spectroscopy (SEM-EDX)) revealed an average composition of 53 % Nb, 45 % P and 2 % Ga with a slight Ga and Nb increase and a P decrease at the edge of the sample (**Fig. 1 e**). Magneto-transport measurements from cryogenic temperatures up to room temperature and with applied magnetic fields up to ±9 T were performed with the current (100 nA) along the [100] axis and the magnetic field varying from the [100] axis (0 °) to the [001] axis (90 °).

At room temperature, the zero-field resistivity of the Ga-doped NbP micro-ribbon is enhanced by a factor of 15 compared to the intrinsic NbP bulk sample. A non-metallic resistivity $\rho$ versus temperature $T$ profile is observed (**Fig. 1 f**). Lowering $E_F$ in $Na_3Bi$ and $Ca_2As_3$ to observe the NMR has previously resulted in a similar $\rho(T)$, which was ascribed to thermal activation of holes across the gapless energy band due to the close vicinity of the Fermi level to the neutrality point.[14, 15]

A quadratic low-field MR (MR = ($\rho(B)$ - $\rho(0)$) / $\rho(B)$ · 100 %, with $\rho(0)$ for zero magnetic field and $\rho(B)$ for an applied magnetic field *B*) and an unsaturated linear high-field magnetoresistance MR are observed in transverse magnetic fields (**Fig. 2 a**). The MR at high fields is attributed to Abrikosov's linear quantum MR,[25] resulting from thermal excitation into the lowest Landau level, which is consistent with the non-metallic $\rho(T)$. Linearity induced by strong disorder[26] is rather unlikely due to the high crystalline quality of our samples. Ultrahigh MR, as observed in intrinsic NbP, is not obtained in our samples (MR(300 K) = 4.46 % to MR(5 K) = 6.42 %), which probably originates from the absence of electron-hole resonance[23] due to a Ga doping-induced Fermi level shift.

Our interpretation is supported by the analysis of strong Shubnikov-de Haas (SdH) oscillations (**Fig. 2 a**), observed below 50 K. After subtracting a smooth background (**Fig. 2 b**), Fast Fourier Transformation (FFT) (**Fig. 2 c**) reveals six fundamental SdH oscillation frequencies - $F_1 = 3.47$ T, $F_2 = 17.37$ T, $F_3 = 24.56$ T, $F_4 = 34.63$ T, $F_5 = 43.08$ T and $F_6 = 71.36$ T - which are correlated with the corresponding electron and hole pockets by performing band structure calculations. NbP exhibits two electron and two hole pockets (E1, E2, H1 and H2) at the Fermi surface.[21] For applied magnetic fields along the [001] axis, hole pockets show one extremal orbit, while electron pockets exhibit 3 extremal orbits (labelled as neck, centre and arm in **Fig. 2 d**). From the best fits of the experimentally observed SdH oscillation frequencies to the $F$ versus $E$ plot obtained from band structure calculations, $F_1$ was identified with the $E_2$ electron pocket, $F_2$ with the $H_2$ hole pocket, $F_3$, $F_5$ and $F_6$ with the $E_1$ electron pocket, and $F_4$ with the $H_1$ hole pocket. Accordingly, the Fermi energy in our Ga-doped NbP micro-ribbons is shifted to $E_F$ (doped) $= + 10$ meV above the intrinsic Fermi level $E_{F0}$ (compare in **Fig. 2 d**). While the W2 points of intrinsic NbP are 5 meV above $E_{F0}$, which prevents these connected W2 points from exhibiting NMR[21], in the doped NbP, the Weyl points 5 meV below $E_F$ are truly separated, which makes them active for chiral anomaly. Given the theoretical position of the Fermi level, oscillation frequencies at 14.02 T and 20.04 T should also be experimentally observable. However, the broad peak at 17.37 T does not allow for sufficient resolution. From the position of the Fermi level, carrier concentrations of $n_h = 1.95 \cdot 10^{19}$ cm$^{-3}$ and $n_e = 4.19 \cdot 10^{19}$ cm$^{-3}$ were calculated such that electron-hole resonance and therefore ultrahigh MR are suppressed. Furthermore, for the most prominent SdH frequency ($F_2 = 17.37$), we obtain a high mobility of $7.1 \cdot 10^5$ cm$^2$/ Vs at 5 K and a relatively low effective mass $m_c = 0.064\ m_0$ from the magnetic field- and temperature-damping of the oscillations.[27]

Tilting the magnetic field parallel to the applied current, we observe a distinct NMR, which we explain via the chiral anomaly, (**Fig. 3 a, b**). Based on the above analysis, the W2 points should be active for the chiral anomaly along the [100] axis. The NMR is very robust against temperature enhancement, as it is observed between 5 and 300 K. Although of the same order of magnitude, in contrast to the positive MR in transverse magnetic fields, the longitudinal NMR at 300 K is enhanced by a factor of 2.9 compared to the NMR at 25 K. We attribute this observation to the ionization of Ga, which increases at elevated temperatures, pushing the Fermi level even closer to the W2 points. Angle-dependent MR measurements (Fig. 3 c) show that the observed NMR is sensitive to the angle ($\phi$) between *B* and *I*, and can be well traced by a $\cos^2(\phi)$-term at low fields, underscoring our assignment to chiral asymmetry. Moreover, at low magnetic fields the inverse of the NMR, the longitudinal positive magnetoconductance MC **Fig. 3 (d)**, is well described by $B^2$- function, in accordance with the prediction of Son and Spivak.[13] At higher fields, however, the longitudinal MC becomes linear. This observation is in agreement with the theoretical description of a transition from a multi-Landau level chiral anomaly to the limit, where only one Landau level is occupied.[25] Conclusively, the longitudinal transition field coincides with the transition to Abrikosov's linear quantum MR in transverse fields. We note that the temperature broadening of the Fermi distribution will activate charge carriers that do not contribute to the NMR. Even at low temperatures likely several band as indicated in **Fig. 2 (d)** will contribute to the transport in parallel to the Weyl cones at zero magnetic field. However, up to the highest temperature of 300 K measured here, we observe that the chiral anomaly seems to dominate the *B*-dependence of *R*, which could be interesting for Weyltronic applications.[19]

In conclusion, we have provided experimental evidence of chiral anomaly in Ga-doped NbP. Our analysis is based on magneto-transport studies of Ga-doped NbP micro-ribbons combined with band structure simulations. The observation of a non-metallic temperature dependence of the resistivity and Abrikosov's linear quantum MR, as well as the evaluation of the observed SdH oscillations, revealed that electron doping shifts the Fermi level in NbP as close as 5 meV below the W2 points, making them active for chiral anomaly. In fact, with parallel magnetic and electric fields, we observe a negative magnetoresistance, whose temperature, magnetic field and angle dependences are consistent with broken chiral symmetry. We believe that this work can pave the way to systematically accessing the unique properties of massless Weyl fermions near the Weyl nodes by material engineering, providing a versatile platform for the development of Weyltronic components.

**Methods:**

   **Crystal growth:** Polycrystalline NbP powder was synthesized by a direct reaction of niobium (Chempur 99.9%) and red phosphorus (Heraeus 99.999%) which were kept in an evacuated fused silica tube ($t$ = 48 h, $T$ = 800 °C). Subsequently, high-quality single crystals of NbP were grown from this microcrystalline powder via a chemical vapour transport reaction in a temperature gradient starting from 850 °C (source) to 950 °C (sink) using iodine (Alfa Aesar 99.998%)[28] with a concentration of 13.5 mgcm$^{-3}$ as a transport agent.

**Crystal structure analysis:** The XRD measurements were performed in a Panlaytical X'Pert four-circle diffractometer using Cu-K$_\alpha$ radiation. The sample was oriented with the help of the Euler cradle in the device. Afterwards, a θ-2θ was taken using a high-resolution optics with a Goebel mirror in the primary beam path and narrow soller slits on the detector side.

**Definition of the micro-ribbon:** From the NbP bulk sample, a 50 μm x 2.46 μm x 526 nm large micro-ribbon was cut with a dual-beam focused ion beam system (FIB) of the type FEI Helios 600i using 30 keV Ga$^+$ ions ($I$ = 65 nA - 80 pA). In order to avoid electrostatic repulsion, the as-cut NbP micro-ribbon was picked up ex-situ with a glass needle and transferred to a glass substrate.

**Compositional analysis of the micro-ribbon:** The compositional analysis was carried out in-situ in the FIB by energy dispersive X-ray analysis (EDX) using 15 keV electrons.

**Definition of the electrical contacts:** The NbP micro-ribbon placed on a 150 μm thick glass substrate was covered with a double layer of photo resist, composed of a layer lift-off resist covered with a layer of positive resist (Micro Chem LOR 3B and maP-1205). The electrical contact structure was defined as shown in **Fig. 1d** using a laser lithography system (Heidelberg Instruments μpg 101). Afterwards, the exposed parts of the photoresist were developed (developer ma-D331, $t = 45$ s, $T = 20$ °C) until an undercut structure was observed at the edges of the developed parts, which ensures a clean removal of the metal layer in the lift-off process. Before the metallization, the NbP contact areas were cleaned by Ar etching ($t = 5$ min, $p = 7.4 \times 10^{-3}$ Torr, flow$_{Ar}$ = 15 sccm and $P = 20$ W) and subsequently 10 nm of Ti and 500 nm of Pt were sputtered without breaking the vacuum. In the lift-off process (Remover 1165, $T = 80$ °C, $t = 1$ h) residual metal parts were removed.

**Transport measurements:** Transport measurements were performed in a cryostatic system (PPMS DynaCool from Quantum Design) equipped with a 9 T electromagnet. The resistance measurements ($I_{ac} = 100$ nA, $f = 189$ Hz) were performed on a rotary sample holder at temperatures between 5 K and 300 K and at a nitrogen pressure of $4 \times 10^{-3}$ mbar.

**Band structure calculations:** The ab initio band structure calculations were performed within the framework of density functional theory (DFT), which are implemented in the Vienna ab initio simulation package.[29] The core electrons were represented by the projector-augmented-wave potential. Generalized gradient approximation (GGA) is employed for the exchange correlation functional. Maximally localized Wannier functions (MLWFs) were used to interpolate the bulk Fermi surface. [30]

**Acknowledgements:**

The authors gratefully acknowledge the work of Tina Sturm and Almut Pöhl defining the NbP micro-ribbon and conducting the compositional analysis and Andy Thomas for helpful discussion. This work was financially supported by the Deutsche Forschungsgemeinschaft DFG via Project DFG-SFB 1143 and DFG-RSF-Project No. (NI 616/22-1) 'Influence of the topological states on the thermoelectric properties of Weyl-Semimetals' and by the ERC Advanced Grant No. (291472) 'Idea Heusler'.

FIGURES

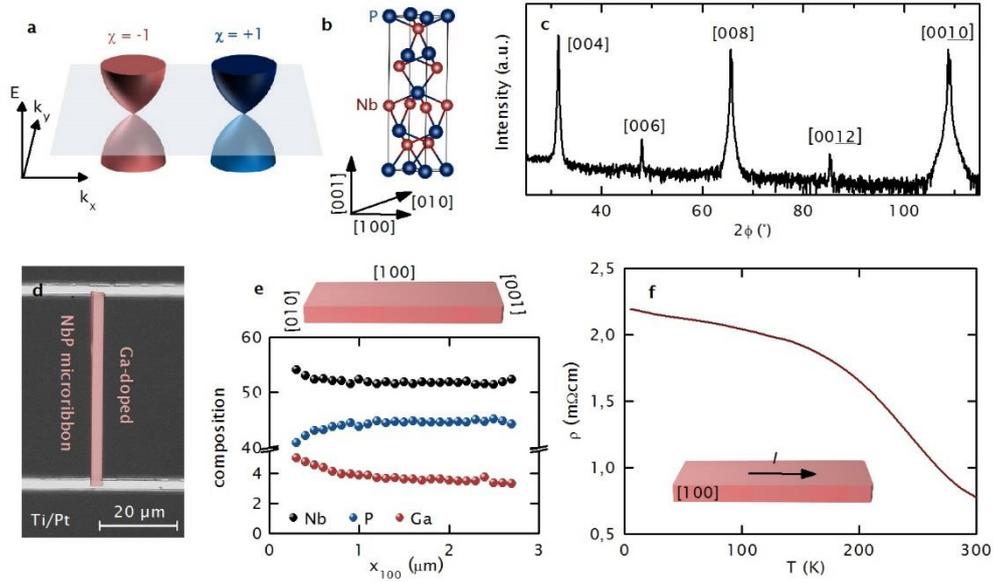

**Figure 1 | Topological semimetal NbP micro-ribbon device.** (**a**) Sketch of a Weyl semimetal, represented as two spatially separated, massless Weyl nodes with distinct chiralities $\chi = -1$ (red cone) and +1 (blue cone). (**b**) The non-centrosymmetric crystal structure in a tetragonal lattice (space group $I4_1md$) of NbP and (**c**) the XRD spectrum of the bulk NbP measured at room temperature. (**d**) Optical micrograph of the NbP micro-ribbon, which has been defined by Ga-FIB. (**e**) SEM-EDX data of the first 3 µm from the left sample edge along the [100] direction of the NbP micro-ribbon reveals an average 53 % Nb, 45 % P and 2 % Ga composition. (**f**) Plot of the resistivity $\rho$ versus temperature $T$.

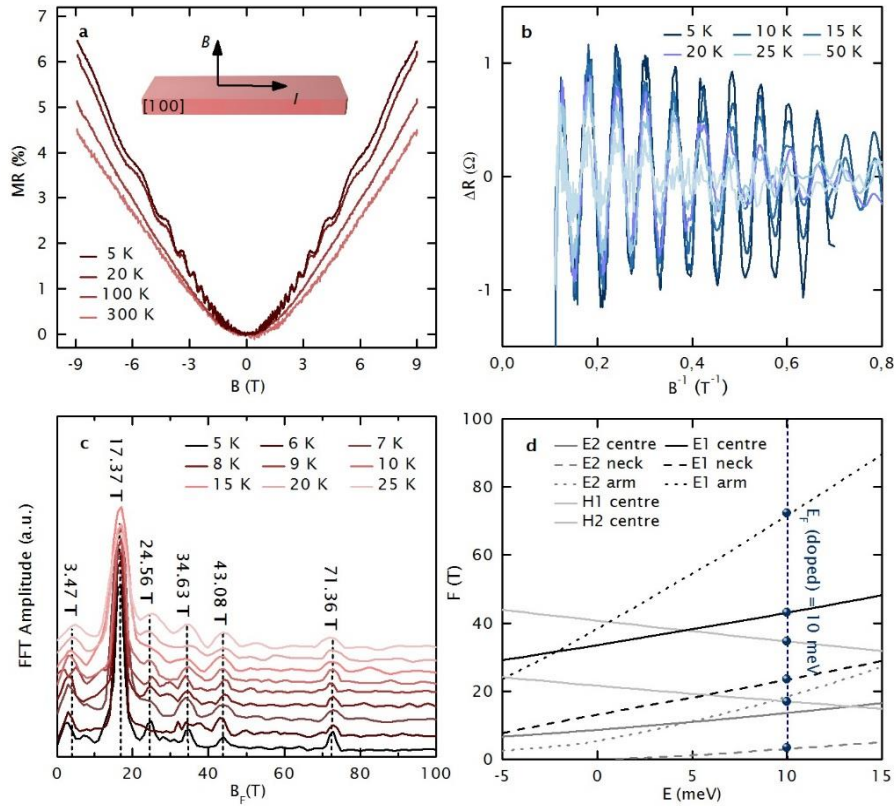

**Figure 2 | Transverse magneto-transport - SdH oscillations and linear high-field magnetoresistance.** (**a**) The temperature-dependent, transverse MR reveals non-saturated linearity at high magnetic fields across the entire $T$ range from 5 K – 300 K and SdH oscillations below 75 K. (**b**) After the subtraction of a non-oscillatory background, the SdH oscillations show a clear periodicity in $B^{-1}$. (**c**) FFT spectra from 5 K to 25 K reveal six fundamental SdH frequencies at $F_1 = 3.47$ T, $F_2 = 17.37$ T, $F_3 = 24.56$ T, $F_4 = 34.63$ T, $F_5 = 43.08$ T and $F_6 = 71.36$ T. (**d**) SdH oscillation frequencies $F$ from *ab initio* simulations are shown as a function of the energy $E$ relative to the intrinsic Fermi level $E_{F0}$. Two electron pockets E1 and E2, which each have three extremal obits (neck, arm and centre), and two hole pockets H1 and H2 with one extremal orbit are resolved. Matching our experimental data (blue dots) to the simulations revealed that $E_F$(doped) is 10 meV above $E_{F0}$ and consequently 5 meV above the W2 nodes.

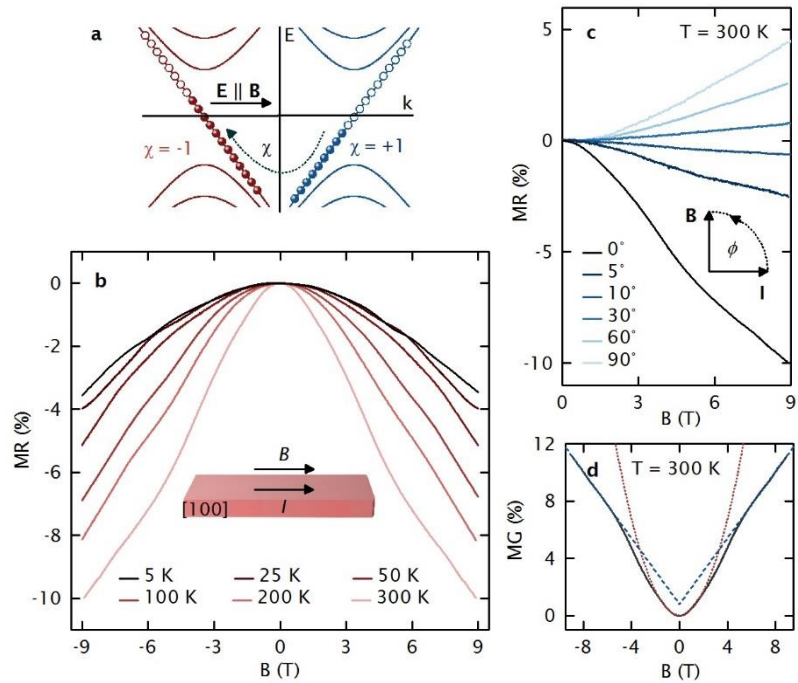

**Figure 3 | Longitudinal magnetotransport – Chiral anomaly-induced negative magnetoresistance.** (**a**) Energy spectrum of left- and right-handed chirality fermions (red and blue, respectively) in parallel applied electric and magnetic fields. In the zeroth Landau level, left-handed particles and right-handed antiparticles have been produced, leading to an additional topological current. (**b**) Temperature dependence of the NMR in parallel magnetic fields. (**c**) Angle-dependent MR at 300 K. (**d**) Positive magneto-conductance at 300 K reveals a parabolic low-field regime that evolves into a linear regime under high magnetic fields.